# Tailoring Magnetic Properties of Zigzag Structured Thin Films via Interface Engineering and Columnar Nano-structuring

Sharanjeet Singh[1], Manisha Priyadarsini[1], Anup Kumar Bera[2], Smritiparna Ghosh[1], Pooja Gupta[3], S.K. Rai[3], Varimalla R. Reddy[1], Velaga Srihari[3], Sarathlal Koyiloth Vayalil[4], Benedikt Sochor[4], Dileep Kumar[1] *

[1] UGC-DAE Consortium for Scientific Research, Khandwa Road, Indore-452001, India
[2] Indian Institute of Science, Bangalore 520016, India
[3] Raja Ramanna Centre for Advanced Technology, Indore 452013, India
[4] Deutsches Elektronen-Synchrotron DESY, Hamburg 22607, Germany
* Corresponding author: dkumar@csr.res.in

**Abstract:** We report the emergence of a novel interface-induced shape anisotropy component in zigzag-structured thin films fabricated via Sequential Oblique Angle Deposition (S-OAD). In this study, we systematically investigate cobalt (Co) and $Co_2FeAl$ (CFA) thin films by varying column length, number of bilayers, and magneto-crystalline anisotropy (MCA) to explore how structural modulation affects magnetic behavior. Using magneto-optical Kerr effect (MOKE) measurements in conjunction with synchrotron-based grazing-incidence small-angle X-ray scattering (GISAXS) and 2D X-ray diffraction (2DXRD), we reveal that the interplay between interface-induced, shape, and crystalline anisotropies allows for a tunable magnetic response, ranging from isotropic to anisotropic behavior. The observed uniaxial magnetic anisotropy (UMA) exceeds that of conventional OAD films, while column merging is effectively suppressed through precise multilayer engineering. Structural analysis confirms that periodic, high-density interfaces at the junctions of oppositely tilted columns are central to this anisotropy control. These findings demonstrate that interface engineering and columnar nanostructuring within zigzag nanostructures offer a powerful route for tailoring magnetic properties in zigzag thin films, enabling their application in next-generation spintronic and magnetic sensor technologies.

## Introduction

The ability to control the magnetic properties of thin films through nano-structuring techniques [1–5] is crucial for advancing applications in data storage [2,6], spintronics [6], and magnetic sensors [7]. In particular, magnetic anisotropy (MA), which refers to the directional dependence of magnetization, is critical in defining the performance of such devices. While uniaxial magnetic anisotropy (UMA) is essential for magnetic sensors [7,8] and memory devices [1,6,7], materials exhibiting isotropic magnetic properties are also indispensable in applications such as magnetic shielding [9,10] and microwave devices [10–12]. Conventional approaches like epitaxial growth offer a pathway to magnetically anisotropic films by exploiting magneto-crystalline anisotropy (MCA) [13]. However, this route often suffers from limitations including material-substrate compatibility, restricted tunability, and modest MCA strength. Consequently, alternative methods that allow broader material choice and finer control over anisotropy are being explored. Among these, Oblique Angle Deposition (OAD) [9,14] has emerged as a powerful technique to engineer magnetic anisotropy via the creation of tilted columnar nanostructures [3–5]. These nanocolumns induce shape anisotropy, which typically aligns the magnetic easy axis either along or perpendicular to the projection of the columns, depending on the oblique angle and film thickness [15–19]. Notably, OAD has proven highly effective in inducing magnetic anisotropy even in polycrystalline and amorphous thin films, which inherently lack MCA due to the absence of long-range order [3,5,20,21]. However, a major limitation of OAD-based anisotropy engineering arises from the merging of columns, which becomes prominent at higher film thicknesses or after thermal annealing [3,5,22,23]. Such morphological coarsening disrupts the distinct columnar structure and leads to a significant reduction in anisotropy strength [1,3,5], thereby compromising the magnetic functionality of the films. To address column merging, some studies have combined OAD with substrate nanopatterning, using periodic ripples to guide column growth and enhance shadowing effects [3,5,20,23–25]. While this strategy helps in maintaining column separation and allows tuning of intercolumnar spacing via the ripple wavelength, it remains effective only up to moderate thicknesses. Moreover, increasing the ripple wavelength can itself weaken magnetic anisotropy due to the modification in shape anisotropy and surface magnetic charges [20,23,26,27].

In this regard, we propose that Sequential Oblique Angle Deposition (S-OAD) [4,9,28–30] can be used as an effective strategy for fabricating multilayer zigzag nanostructures that maintain distinct columnar morphology throughout the film without significant column merging [31–34]. The S-OAD approach is known to promote higher porosity compared to conventional OAD, which can effectively preserve inter-column separation even at larger film thicknesses, up to 200-300 nm, provided the individual layer thickness remains less than ~40 nm [4,31,32,34]. This increased porosity is governed by the no. of bilayers, the thickness of individual layers, and the alternating deposition angles, all of which can be precisely tuned to control the spatial arrangement and anisotropic properties of the columns [9,31,34]. The zigzag architectures have demonstrated significant improvements in surface area, mechanical robustness, and structural functionality across various material systems, including metal nitrides, oxides, and glasses [9,14,35–39]. While zigzag nanostructures have been explored in electronic [35,39], photonic [29,36], and catalytic systems [37,38], their integration into magnetic thin films remains limited. A few previous studies have attempted to implement zigzag nanostructuring in magnetic systems [4,33,40–42], but with significant limitations. For instance, some employed very thick individual layers (>100 nm) without addressing the resulting column merging, as their primary focus was not on systematically analyzing in-plane magnetic anisotropy [40,41]. Other studies, though successful in inducing in-plane MA using S-OAD while minimizing column merging [4,33,42], overlooked the presence of finite coercivity along the hard axis, a feature suggestive of an additional anisotropy component orthogonal to the column projection. In our earlier work on cobalt zigzag films

[28], we observed such unusual isotropic magnetic behavior, hinting at the emergence of a novel anisotropy component arising from the structural features of zigzag nanostructuring, which competes with conventional shape and crystalline anisotropies. While conventional OAD has been widely studied for optimizing uniaxial magnetic behavior, the potential of S-OAD-induced zigzag nanostructures for tailoring magnetic properties remains largely underexplored. Addressing this research gap is crucial for developing films via S-OAD that can finely tune both anisotropic and isotropic magnetic behavior to meet specific application requirements.

In an effort to bridge this gap, the present study systematically investigates the influence of column length, number of bilayers, and MCA in cobalt (Co) and $Co_2FeAl$ (CFA) thin films fabricated using S-OAD. By combining magneto-optical Kerr effect (MOKE) measurements with synchrotron-based grazing-incidence small-angle X-ray scattering (GISAXS) and 2D X-ray diffraction (2DXRD), we explore how the interplay between shape, crystalline, and interface-induced anisotropies governs the magnetic response of these zigzag films. Through precise multilayer engineering, we have successfully synergized the effects of dipolar interactions and shape anisotropy along a single, well-defined direction, while simultaneously preventing column merging. The observed strength of UMA is even higher than conventional OAD-induced UMA. It has been demonstrated that through careful interface engineering and columnar nanostructuring within zigzag nanostructured films, the magnetic behavior can be finely tuned, offering a viable pathway for designing thin films with customized magnetic properties. Overall, this work advances the understanding of S-OAD grown films and highlights zigzag nanostructuring as a powerful strategy for controlling magnetization in thin films for next-generation spintronic and magnetic sensor applications.

## Experimental

A series of cobalt (Co) zigzag thin films with varying bilayer numbers (3, 4, 5, and 7 bilayers) were deposited on Si (100) substrates (from Sigma-Aldrich) using electron-beam evaporation. The Sequential Oblique Angle Deposition (S-OAD) method was employed to obtain zigzag films by alternating the deposition angle between +70° and -70° relative to the substrate normal, producing a multilayered zigzag nanostructure. Figure 1 (a) illustrates the deposition geometry used to create a bilayer, which was repeated to form multilayered structures. Each bilayer [$Co_{\alpha=70°}$ (3.5 nm)/$Co_{\alpha=-70°}$ (3.5 nm)] consists of two individual layers of thickness 3.5 nm deposited at alternate angles, resulting in a bilayer of thickness ~7 nm. Additionally, a conventional OAD [$Co_{\alpha=70°}$ (50 nm)] film, a single bilayer zigzag [$Co_{\alpha=70°}$ (25 nm)/$Co_{\alpha=-70°}$ (25 nm)] sample, and a 3-bilayer $Co_2FeAl$ (CFA) [$CFA_{\alpha=70°}$ (3.5 nm)/$CFA_{\alpha=-70°}$ (3.5 nm)]$_3$ zigzag thin film were fabricated to examine the influence of layer thickness (or columnar length) and magneto-crystalline anisotropy (MCA) on the magnetic behavior. The base pressure in the deposition chamber was ~ $2 \times 10^{-8}$ mbar, and the working pressure during deposition was maintained at ~ $4 \times 10^{-7}$ mbar, with a deposition rate of 0.3 Å/s. A calibrated quartz crystal monitor (Telemark deposition controller, model 860) was used to precisely track the film thickness under oblique angle deposition conditions.

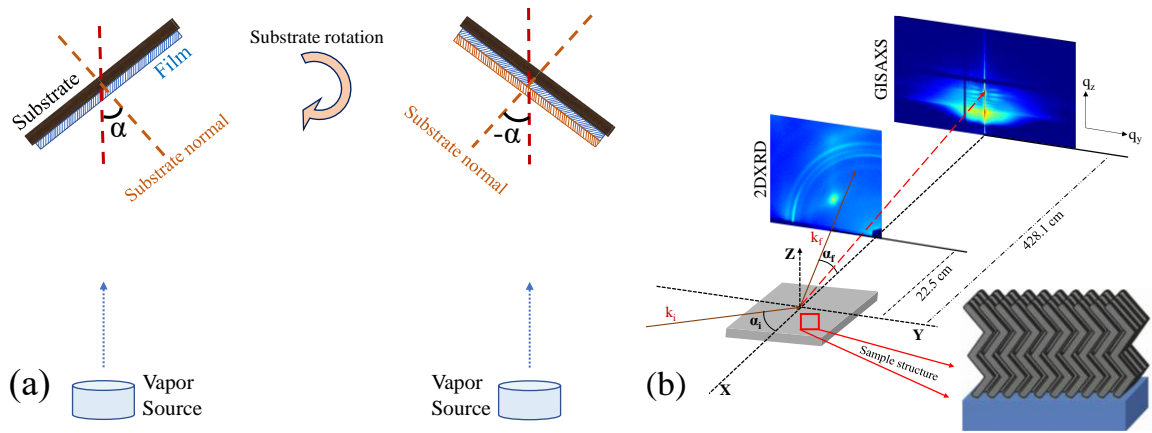

**Figure 1.** (a) shows the schematic of the deposition geometry of a bilayer. (b) shows the schematic of the GISAXS and 2DXRD geometry along with the sample structure.

The magnetic properties were characterized using Magneto-Optical Kerr Effect (MOKE) measurements in the longitudinal geometry, where hysteresis loops were recorded at multiple azimuthal angles ($\phi$) with respect to the column projection. To gain insights into the structural and morphological properties, synchrotron-based grazing-incidence small-angle X-ray scattering (GISAXS) measurements were conducted on the 7-bilayer zigzag, 1-bilayer zigzag, and conventional OAD samples. The experiments were performed at beamline P03 [43] of PETRA III, DESY, Germany, using X-rays of 11.87 keV ($\lambda$ = 1.044 Å). The scattered intensity was recorded with a PILATUS 2M (Dectris Ltd.) detector (pixel size of 172 × 172 μm²) at a sample-to-detector distance of 4281 mm. To investigate the evolution of crystalline structure with increasing bilayer numbers, grazing-incidence X-ray diffraction (GIXRD) measurements were performed on the 3, 4, and 5-bilayer zigzag samples at the Engineering Applications beamline (BL-02) [44] of Indus-2 synchrotron, RRCAT, India, using X-rays of 17 keV ($\lambda$ = 0.7283 Å). To investigate crystallographic texture, 2D X-ray diffraction (2DXRD) measurements were performed on the 7-bilayer zigzag and conventional OAD samples of Co and a 3-bilayer zigzag sample of CFA, at the AD/ED-XRD beamline (BL-11) [45] of Indus-2 synchrotron, RRCAT, India, using X-rays of 17 keV ($\lambda$ = 0.7283 Å). A Mar345 area detector (marXperts GmbH) with a 100 × 100 μm² pixel size was used at a sample-to-detector distance of 225 mm. GISAXS and 2DXRD measurements were performed at a grazing incidence angle of 0.4°, with data collected in two orientations: (i) along the column projection ($\phi$ = 0°), (ii) perpendicular to the column projection ($\phi$ = 90°). The schematic of the GISAXS and 2DXRD measurement geometries is shown in Fig. 1 (b) along with a schematic of the sample structure. The analysis for GISAXS and 2DXRD images was performed using DPDAK software [46], and the extracted structural and morphological parameters were correlated with the observed magnetic anisotropy trends.

# Results and discussion

**a. Magnetic studies:**

The evolution of magnetic properties of zigzag thin films with varying bilayer numbers was investigated using MOKE measurements. Figure 2 (a-d) presents the hysteresis loops taken along ($\phi$ = 0°) and perpendicular ($\phi$ = 90°) to the projection of columns for different samples. The corresponding polar plots of normalized remanence (Mr/Ms) as a function of the in-plane azimuthal angle, shown in Fig. 2 (e-h), provide insights into the systematic transition of the easy axis with increasing bilayer count. A 3D schematic of the sample structures is displayed alongside the respective hysteresis loops.

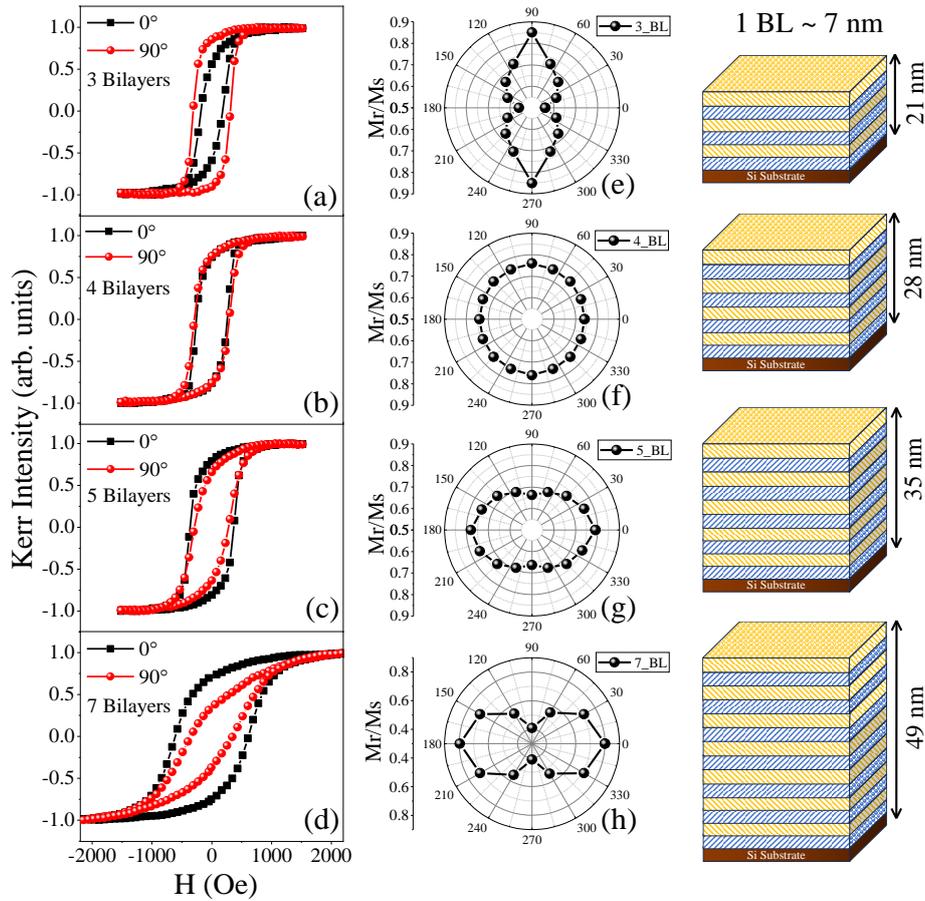

**Figure 2.** (a-d) MOKE loops of different numbers of bilayers. (e-h) shows the corresponding polar plots of normalized remanence. The schematic of the sample structure is shown alongside.

The 3-bilayer (3-BL) sample exhibits a weak uniaxial magnetic anisotropy (UMA), with the easy axis oriented perpendicular to the column projection ($\phi$ = 90°). Upon increasing the bilayer count to 4 (4-BL), the film develops an isotropic magnetic response, where no clear preferential easy axis is observed. For the 5-bilayer (5-BL) sample, the easy axis gradually shifts towards the column projection ($\phi$ = 0°), and this trend becomes more pronounced in the 7-bilayer (7-BL) sample, indicating a progressive dominance of the column-projection-favored anisotropy.

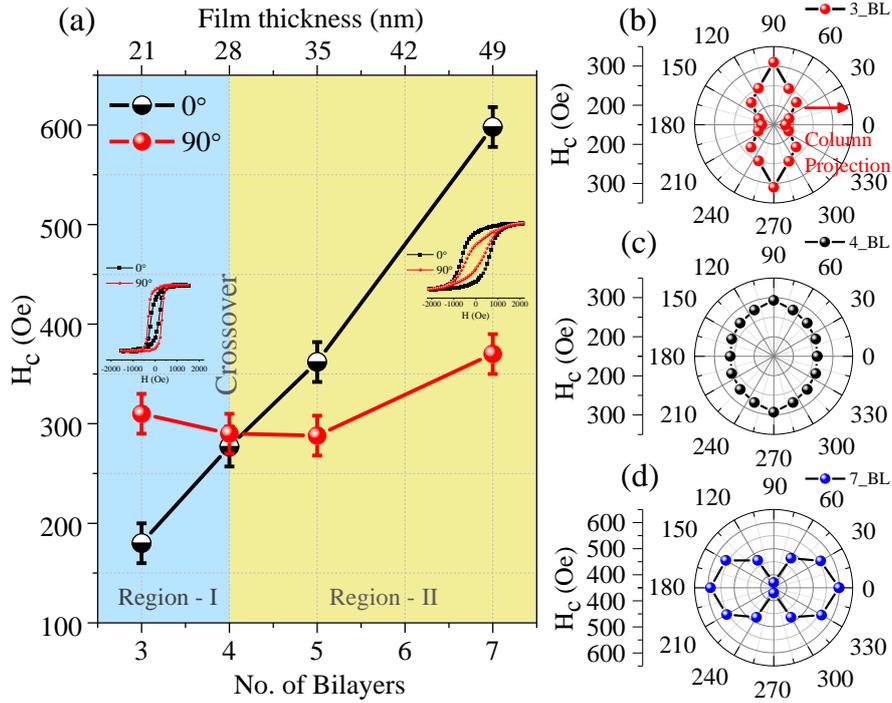

**Figure 3.** (a) Variation of coercivity along ($\phi = 0°$) and perpendicular ($\phi = 90°$) to the column projection with increasing number of bilayers. (b-d) shows the polar plots of coercivity of 3, 4, and 7 bilayer samples.

Figure 3 (a) presents the variation of coercivity ($H_c$) with the number of bilayers for $\phi = 0°$ and $\phi = 90°$, revealing a crossover in anisotropy behavior as the bilayer count increases. In Region I (3 bilayers), the coercivity for $\phi = 90°$ is greater than for $\phi = 0°$, whereas it becomes comparable as the bilayer count increases to 4. In Region II (5,7 bilayers), the coercivity along $\phi = 0°$ surpasses that at $\phi = 90°$, and increases progressively with bilayer count. To further investigate this crossover, polar plots of coercivity for the 3, 4, and 7-bilayer samples are presented in Fig. 3 (b-d). The 3-bilayer sample displays a characteristic "figure-eight" shape [3] elongated along $\phi = 90°$, confirming a preferential easy axis perpendicular to the column projection. At 4 bilayers, the nearly circular profile suggests an isotropic response. However, as the bilayer count increases to 7, the polar plot again exhibits an eight-like shape, but now more elongated along $\phi = 0°$, indicating the re-emergence of UMA with an easy axis shifted along the column projection. A clear anisotropy crossover is observed here, marking the transition of UMA's easy axis along the projection of columns from its perpendicular direction by gradually passing through the isotropic state. This kind of magnetic behaviour is uncommon in cases of OAD films, thus to further investigate this evolution of anisotropy, two additional samples were examined, including a conventional OAD film and a 1-bilayer (25+25 nm) zigzag sample to measure the influence of number of bilayers and the column length (or layer thickness) on their magnetic properties.

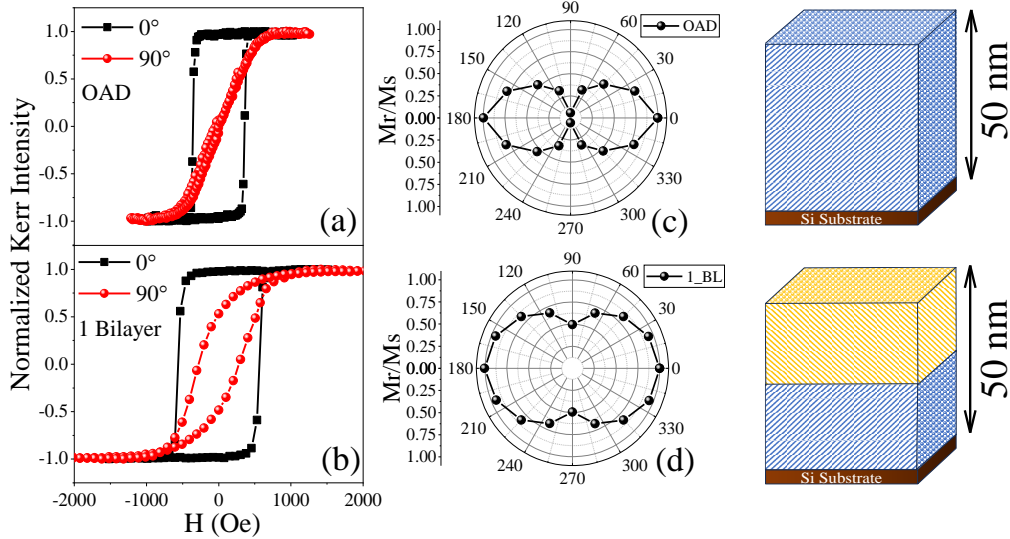

**Figure 4.** (a, b) shows the MOKE loops of conventional OAD and 1 bilayer samples. (c, d) shows the corresponding polar plots of normalized remanence. The schematic of the sample structure is shown alongside.

Figure 4 presents the MOKE hysteresis loops, polar plots of normalized remanence, and 3D depictions of sample structures for these films. The conventional OAD sample exhibits strong UMA, with the easy axis aligned along the column projection ($\phi = 0°$), a behavior characteristic of obliquely deposited thin films due to the shape anisotropy induced by tilted columnar growth [17,18]. Similarly, the 1-bilayer (1-BL) sample displays pronounced anisotropic behavior, favoring the column projection. However, a key distinction emerges in the hard-axis ($\phi = 90°$) response. While the conventional OAD film shows an almost closed hysteresis loop along $\phi = 90°$, the 1-bilayer sample exhibits an open loop, suggesting the presence of an additional factor influencing the magnetic behavior, potentially favoring alignment perpendicular to the column projection.

**b. GISAXS Studies:**

To explore this anisotropy crossover behaviour and its correlation with morphology and structure, we have performed GISAXS measurements at P03 beamline, PETRA-III, DESY. Figure 7 (a-f) shows the GISAXS images of the 7-bilayer, 1-bilayer zigzag, and conventional OAD cobalt thin films collected with the X-ray beam aligned along ($\phi = 0°$) and perpendicular ($\phi = 90°$) to the projection of the column. For $\phi = 0°$, the images of the 1-bilayer and conventional OAD samples are nearly the same, while the 7-BL sample shows some horizontal lines. On the other hand, the images taken along $\phi = 90°$ exhibit clear differences from each other. 7-bilayer sample image shows distinct horizontal streaks replicating in the vertical direction and positioned symmetrically on either side of the scattering rod. This indicates the well-defined lateral and vertical periodicities along the film thickness introduced by the alternating tilt of columns in successive layers [28]. 1-bilayer sample shows distinct broad plumes symmetrically positioned on either side of the specular rod, introduced by the alternating tilt of columns in successive layers. The plume on the right side (marked by a red dotted curve) of the scattering rod is less intense as compared to the plume present on the left. It is because the top layer is 25 nm thick, and most of the scattering takes place from this top layer, thus, X-rays scattered from the bottom layer of 25 nm are of lower intensity. As expected, the GISAXS image for conventional OAD lacks such symmetric features and instead displays an asymmetric intensity distribution having the plume only on the left side of the specular rod. It is due to the uniform tilt of the columns only in one direction [3,18]. The angle of tilted columns $(β)$ relative to the surface normal is found to be 47° ± 3° for the OAD sample, whereas this tilt angle increases to 53° ± 3° for the 1-BL sample.

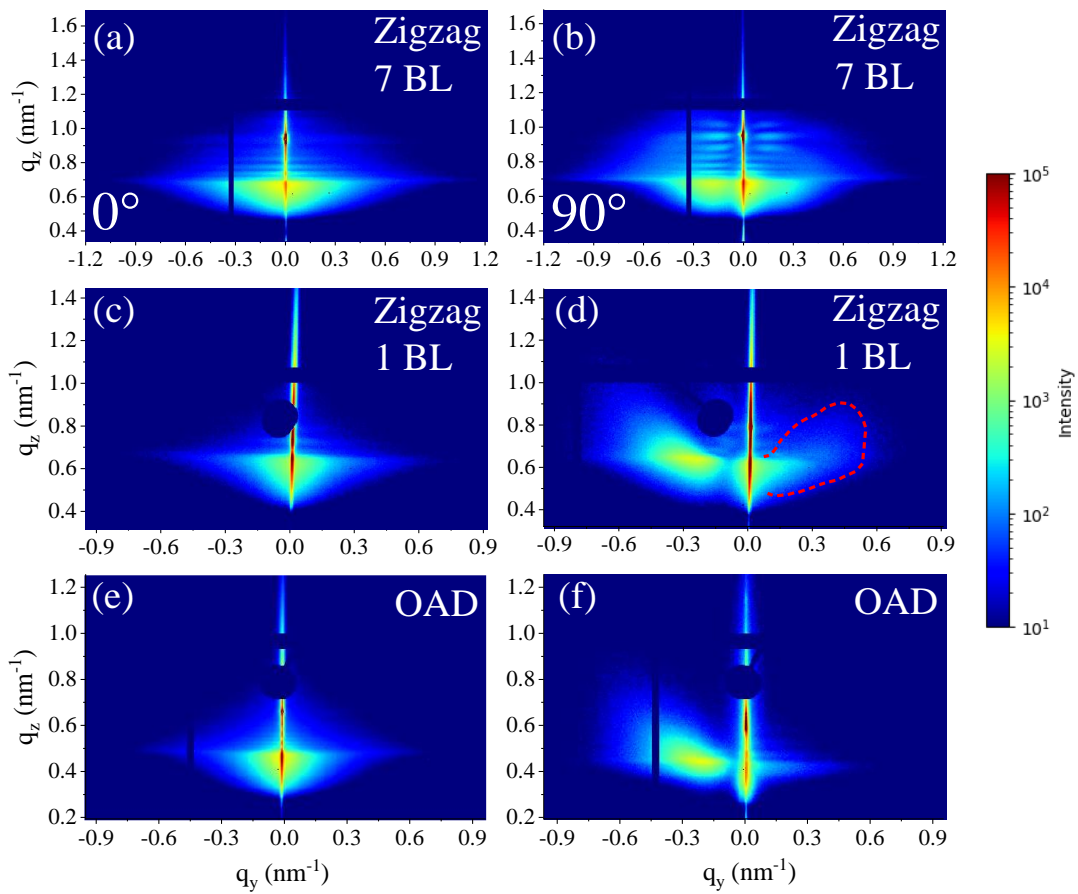

**Figure 5**. GISAXS images of the (a, b) 7 bilayer zigzag, (c, d) 1 bilayer zigzag, and (e, f) conventional OAD samples for $\phi = 0°$ and $\phi = 90°$.

To obtain the critical insights into the lateral scattering behavior of the zigzag multilayer and uniformly tilted columnar structures, the horizontal line integration curves were extracted from the GISAXS patterns of the 7-bilayer and OAD samples for $\phi = 0°$ and $\phi = 90°$. The extracted in-plane line profiles are shown in Fig. 6 (a) for the zigzag and in the inset for the OAD sample. For the 7-bilayer zigzag sample, the integration profile along $\phi = 90°$ exhibits two broad peaks symmetrically positioned on either side of the scattering rod. The appearance of two broad peaks in the zigzag films is a direct result of the zigzag deposition sequence, where the flux alternates between two oblique angles, generating two sets of tilted columns oriented in opposite directions [28]. These peaks arise due to the alternating tilt of the columns in successive layers, where each layer scatters to opposite sides, creating a distinct lateral periodicity. On the other hand, the line profile for the OAD sample shows only a single asymmetric peak due to the uniform tilt of the columns in one direction [3,18]. Such features are absent in the corresponding integration curve for ϕ = 0° for 7-bilayer zigzag and OAD samples, where no clear lateral peaks are observed, as the orientation of columns lies in the vertical scattering plane.

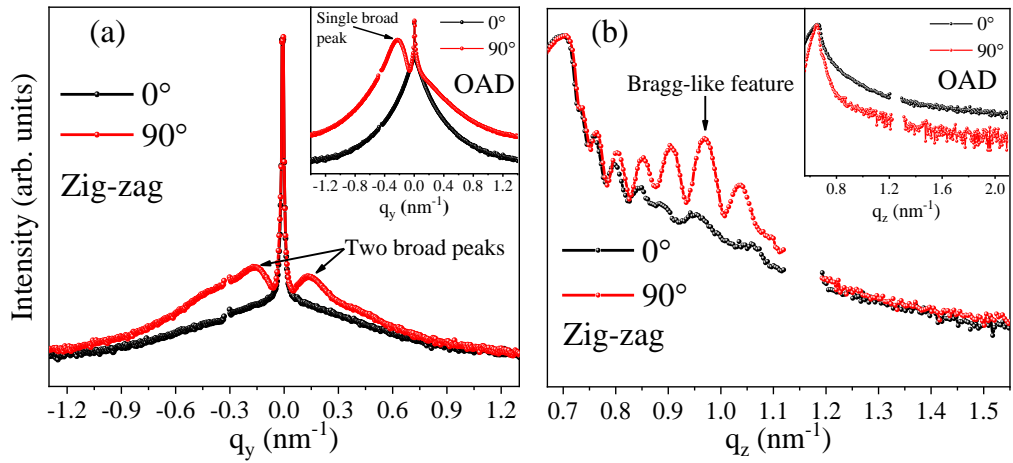

**Figure 6.** (a) Horizontal line profile of the 7-BL sample along ϕ = 0° and ϕ = 90°. Inset shows the horizontal line profile of the OAD sample. (b) vertical line profile of the 7-BL sample ϕ = 0° and ϕ = 90°. Inset shows the vertical line profile of the OAD sample.

To extract the intensity (I) vs. out-of-plane momentum transfer vector ($q_z$) profiles, a vertical cut (indicated as the white dashed line in Fig. 5 b) was made slightly left of the specular rod. The extracted profiles for 7-bilayer zigzag and conventional OAD samples are plotted in Fig. 6 (b) and their insets, respectively. The vertical line profile of the 7-bilayer zigzag sample along ϕ = 90° exhibits intensity oscillations resembling Kiessig-like fringes and a Bragg-like feature, whereas both of these features are missing in the line profile of the OAD sample. The Kiessig-like oscillations, along with a distinct Bragg-like feature, provide critical insights into the periodicity and nanoscale structural organization of the multilayer zigzag architecture. The occurrence of Kiessig-like oscillations is a measure of conformal growth of the film due to a periodic and alternate tilt of the column in the successive layers of the 7-bilayer zigzag sample. The origin of the Bragg-like peak lies in the periodic electron density contrasts introduced by the S-OAD process, which creates alternating layers of tilted columns oriented in opposite directions, giving rise to the dense interfaces at their junction [28]. Each interface between the layers contributes to a periodic modulation in electron density, resulting in coherent scattering that produces Bragg-like features. The position of the Bragg peak along the vertical momentum transfer vector ($q_z$ ~ 0.96 ± 0.01 nm$^{-1}$) directly corresponds to the periodicity length (d ~ 6.5 ± 0.7 nm), which can be estimated using Bragg's law for small angles:

$$d = 2\pi/q_z$$

The periodicity length revealed by the Bragg-like peak closely matches the bilayer thickness (7 ± 1 nm), which confirms the consistent stacking of zigzag layers matching the intended deposition parameters. Thus, each bilayer in the zigzag arrangement contributes to the overall periodicity, creating a repeating structure with high-density interfaces across the film thickness. On the other hand, the absence of Bragg-like features in the line profile of the conventional OAD sample indicates that the film has a uniform density profile without any interfaces, corresponding to the regular columnar arrangement formed by the OAD.

### c. Structural characterization through GIXRD and 2DXRD:

The structural evolution of zigzag films with increasing bilayer number was investigated using GIXRD measurements. A full 2θ scan was initially performed (data not shown), followed by a focused investigation of the q-range covering the (100), (002), and (101) reflections of hexagonal close-packed (hcp) Co, which appeared as the dominant peaks in the full scan. The absence of diffraction peaks corresponding to cobalt oxide phases in the full 2θ scan confirms the pristine nature of the films, indicating they are largely free from oxidation. Figure 7 (a-c) presents the out-of-plane XRD patterns, which show a systematic decrease in the relative intensity of the (002) peak with increasing bilayer number, indicating possible changes in preferred orientation or texture. To further investigate the texture associated with the (002) planes, in-plane GIXRD measurements were performed at $\phi = 0°$ and $\phi = 90°$, aligning the momentum transfer vector (q) perpendicular and parallel to the projected direction of the tilted columns, respectively. As shown in Fig. 7 (g-i), the relative intensity of the (002) peak increases along the projection of the columns with increasing bilayer number, while remaining nearly constant in its perpendicular direction [Fig. 7 (d-f)]. This directional trend suggests the progressive development of in-plane texturing of the (002) planes along the columnar projection with increasing bilayer number. To quantitatively assess the structural evolution, peak fitting of the XRD patterns was carried out. The extracted peak position values for all three planes remain nearly constant for out-of-plane and in-plane configurations across all bilayer samples, suggesting the absence of measurable strain. For a more precise evaluation of the (002) texturing behavior, the normalized area under the (002) peaks was calculated and is displayed alongside the corresponding peaks in Fig. 7 (d-i). Along the columnar projection, the area under the (002) peak increases significantly, from 0.42 for the 3-bilayer sample to 0.60 for the 5-bilayer sample, while it remains nearly constant (~0.38) in the perpendicular direction. The relative enhancement of the (002) peak area along the column projection compared to the perpendicular direction is approximately 11% for 3 bilayers, 20% for 4 bilayers, and 50% for 5 bilayers. This quantitative trend provides strong evidence for the progressive development of in-plane crystallographic texture along the direction of columnar projection with increasing bilayer number. In addition to texturing, the crystallite size was found to be anisotropic. Specifically, the full width at half maximum (FWHM) of the (002) peak along the projection of the columns is consistently lower than that in the perpendicular direction, indicating a larger crystallite size along the columnar axis. Moreover, this anisotropy becomes more pronounced with increasing bilayer number. The FWHM ratio (perpendicular to parallel direction) increases from 1.13 for 3 bilayers to 1.16 for 4 bilayers and further to 1.57 for 5 bilayers, suggesting enhanced crystallite growth along the columnar projection.

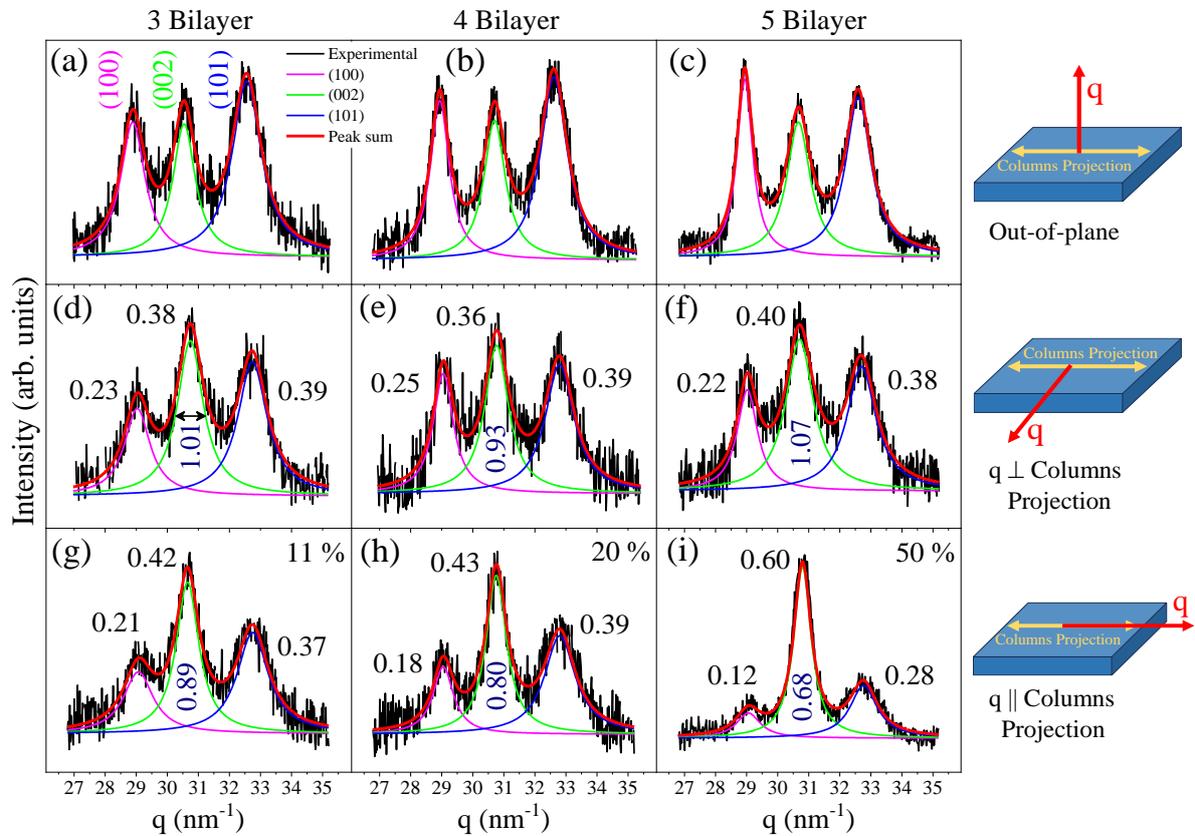

**Figure 7.** (a-c) Out-of-plane and (d-i) in-plane GIXRD patterns of the 3, 4, and 5 bilayer samples. The respective measurement geometry is shown alongside the GIXRD patterns. The normalized area under each diffraction peak is indicated adjacent to the corresponding reflection, while the FWHM of the (002) peak is shown within the peak profile. The error in the normalized area is ±0.03, and in the FWHM is ±0.05.

To evaluate the full orientation distribution of crystallographic planes, 2DXRD measurements were conducted on the 7-bilayer zigzag and conventional OAD films with the X-ray beam incident along $\phi = 0°$ and $\phi = 90°$. As shown in Fig. 8, both samples exhibit diffraction arcs from (100), (002), and (101) planes of hcp Co [5,47]. For $\phi = 0°$, the intensity of all three arcs remains uniformly distributed across the azimuthal range ($\chi \approx 0°-180°$) for both of the samples. In contrast, for $\phi = 90°$, a pronounced azimuthal intensity variation is observed in the (002) arc for both samples, revealing preferential orientation (texturing) of the (002) planes along the projection of the columns. However, the nature of this (002) texturing differs significantly between the two samples. The 7-BL zigzag film exhibits a sharp and symmetric intensity enhancement near the in-plane direction, confined to a narrow azimuthal spread ($\chi \approx 0°-15°$) on both sides of the detector. This reflects a strong, well-aligned in-plane texture arising from the periodic, alternating tilt of columns in successive layers. On the other hand, the OAD sample shows a relatively weaker (002) arc intensity, limited to one side (right) of the detector and spread over a broader azimuthal range ($\chi \approx 0°-60°$). This asymmetric intensity distribution indicates a unidirectional column tilt and texturing that predominantly occurs along a single growth direction. Thus, while both films show (002) texturing, the zigzag structure promotes more symmetric and sharply confined strong in-plane texture due to its alternating growth geometry, in contrast to the broader, asymmetric texture of the OAD sample due to the uniform tilt of columns.

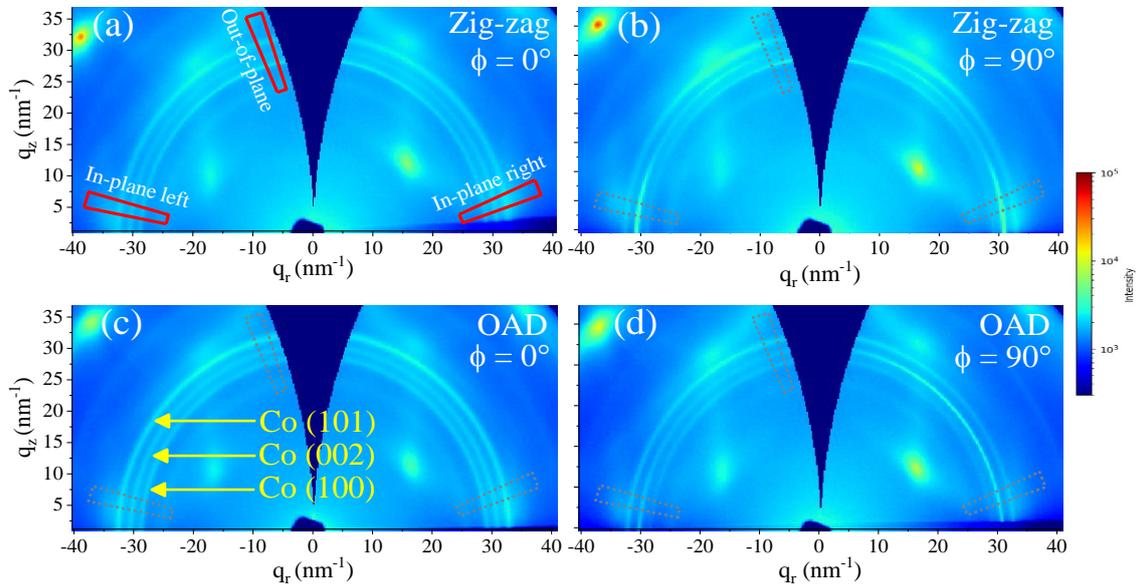

**Figure 8.** Shows the missing wedge corrected 2DXRD images of the 7-bilayer and OAD sample while keeping the X-ray direction along ($\phi = 0°$) and perpendicular ($\phi = 90°$) to the column projection.

For quantitative analysis, in-plane data were extracted from both left and right sides of the 2DXRD patterns of 7-bilayer zigzag and OAD samples using DPDAK software. The positions of radial cuts are marked by red sections in Fig. 8 (a). The XRD patterns were fitted, as shown in Fig. 9, to extract quantitative information including the peak positions and area under the (100), (002), and (101) reflections for both zigzag and OAD samples. The peak positions of each plane were found to be consistent along and perpendicular to the column projection for both samples, indicating the absence of any in-plane strain in both samples.

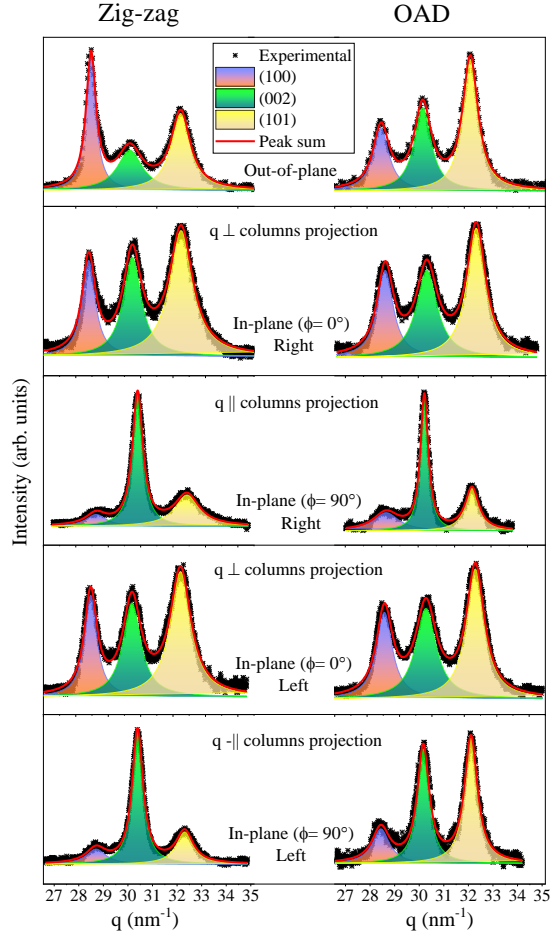

Figure 9. The fitted out-of-plane and in-plane 2DXRD plots for $\phi = 0°$ and $\phi = 90°$ extracted from the 2DXRD images of 7-bilayer zigzag and OAD samples.

The normalized peak areas for zigzag and OAD samples are summarized in Table I for all three reflections. A point-wise comparison highlighting the key structural differences between the 7-bilayer zigzag and OAD samples is presented below.

- In the out-of-plane XRD data, the normalized area of the (002) peak is lower for the zigzag sample than for the OAD sample, indicating that zigzag nanostructuring inhibits the out-of-plane alignment of the (002) planes. Notably, the zigzag sample shows a relatively higher intensity of the (100) peak, which is perpendicular to (002), suggesting a stronger in-plane orientation of the (002) planes compared to the OAD structure.

- The in-plane data extracted from the right side shows that both samples exhibit an enhanced area under the (002) peak along the projection of columns. However, this anisotropy is more pronounced in the zigzag sample, where the (002) peak area is ~93% higher along the column projection than in the perpendicular direction, compared to only ~56% in the OAD sample.

- Compared to the in-plane data extracted from the right side, the left-side data reveals a striking difference in texturing behavior. For the OAD sample, the (002) texturing decreases sharply, with the area difference dropping from ~56% to just ~15%. In contrast, the zigzag sample maintains strong (002) texturing along the column projection, with the area enhancement increasing to ~110%. This highlights that zigzag nanostructuring induces robust and persistent in-plane texturing of the (002) planes along the columnar projection, whereas the OAD structure produces relatively weaker texturing that is confined to a specific direction, i.e., along the tilt direction of columns. Therefore, one must take care while performing XRD on OAD samples, as it is crucial to align the momentum transfer vector (q) along the tilt direction of the columns to accurately capture the extent of texturing.

Table I. The fitting results for normalized area *(A)* of (100), (002), and (101) planes along various out-of-plane and in-plane directions for zigzag and OAD samples. The error bar derived for the normalized area *(A)* from the least-squares fitting of the XRD data is ±0.02.

| XRD (Geometry) | | Co (100) | | Co (002) | | Co (101) | |
|---|---|---|---|---|---|---|---|
| | | Area Zig-zag | Area OAD | Area Zig-zag | Area OAD | Area Zig-zag | Area OAD |
| Out-of-plane | q (normal) Film plane | 0.35 | 0.22 | 0.26 | 0.32 | 0.39 | 0.46 |
| In-plane Right | q ⊥ columns projection | 0.21 | 0.25 | 0.30 | 0.32 | 0.49 | 0.43 |
| | q ∥ columns projection | 0.10 | 0.22 | 0.58 | 0.50 | 0.32 | 0.28 |
| In-plane left | q ⊥ columns projection | 0.23 | 0.25 | 0.30 | 0.33 | 0.47 | 0.42 |
| | q –(∥) columns projection | 0.13 | 0.17 | 0.63 | 0.38 | 0.24 | 0.45 |

### d. Discussion:

In systems where anisotropy crossover is observed [48,49], it typically results from the competition between orthogonal anisotropy components. In the present study, such a crossover is observed as the number of bilayers increases from 3 to 7. The 3-bilayer system exhibits weak UMA with the easy axis oriented perpendicular to the projection of the columns, while the 7-bilayer system shows a shift of the easy axis along the column projection. Interestingly, the 4 and 5-bilayer systems display nearly isotropic magnetic behavior, marking a transition regime between the two orthogonal anisotropy orientations. This kind of behavior is uncommon in the case of OAD films, which are known to exhibit anisotropic behavior. XRD analysis reveals that the (002) planes (c-axis) of Co films are preferentially aligned along the projection of the columns, and this texturing becomes more pronounced with increasing bilayer count. Additionally, the crystallite size is anisotropic, with larger crystallites oriented along the column projection. This anisotropy in grain size may promote local magnetic alignment by reducing domain wall pinning and enhancing exchange interactions. However, despite the presence of both texture-induced MCA and crystallite size effects favoring alignment along the column projection, the zigzag films do not exhibit a strong UMA or a square-shaped hysteresis loop in that direction. This suggests the presence of a competing anisotropy component that tends to align the magnetization perpendicular to the column projection. In conventional OAD films, tilted columnar structures form via geometric shadowing, where initial nucleation sites block incoming adatoms, leading to anisotropic growth. For film thicknesses above ~15 nm, these columns typically induce shape anisotropy favoring magnetization along the column projection [18,50]. Given that the individual layer thickness is small (~3.5 nm) in zigzag films, the columnar shape-induced anisotropy is minimal due to the limited aspect ratio. It is not expected to contribute significantly in either direction. However, it has been observed in the case of conventional OAD films that, at low thickness, generally below ~10 nm, the easy axis tends to align in the perpendicular direction of the column projection [17–19,50,51]. In this regime, the weak aspect ratio suppresses shape anisotropy, and the magnetic behavior is dominated by dipolar interactions between columns, which favors the easy axis in the perpendicular direction due to the lower intercolumnar separation in this direction [17–19,50,51]. However, the overall strength of MA in this low-thickness regime remains relatively modest, as it originates primarily from these weak dipolar interactions rather than from the shape or intrinsic structural anisotropy. Therefore, dipolar interactions alone are insufficient to overcome the stronger MCA and crystallite-driven anisotropies observed in our system. GISAXS analysis reveals that zigzag nanostructuring via S-OAD leads to the formation of high-density interfaces at the junctions of oppositely tilted columns. It is a well-established fact that OAD leads to the formation of tilted columns with an anisotropic intercolumnar separation [3,52,53]. The columns remain well separated from each other along their tilt direction, whereas in the perpendicular direction, the columns grow broader and remain nearly connected [3,52,53]. This structural asymmetry and interconnection of columns perpendicular to the projection of columns can lead to the formation of rod-like structures at the dense interfaces, with their long axis oriented perpendicular to the projection of columns. These dense, elongated interface regions can act as magnetic entities that contribute an additional shape anisotropy component perpendicular to the projection of the columns. The periodic nature of the interfaces formed in zigzag structures, combined with the small individual layer thickness (~3.5 nm), makes the interface-induced shape anisotropy particularly significant. These interfaces are spaced ~3.5 nm apart, which is less than the exchange length of ~5 nm [19], allowing strong exchange coupling between adjacent layers. As a result, the interfaces can collectively influence the magnetization reversal of the columnar segments sandwiched between them. This entire assembly of tilted columns bounded by interfaces may effectively behave as a rectangular magnetic plate, with its long axis extending a few microns perpendicular to the column projection, and its width limited to just a few nanometers, limited by the typical column width. Such high-aspect-ratio structures can generate an additional shape anisotropy component in the direction perpendicular to the column projection. Meanwhile, the strong texturing-induced magneto-crystalline anisotropy and anisotropic crystallite size favor magnetization along the column projection. The competition between these orthogonal anisotropy components, interface-induced and dipolar versus MCA and crystallite-driven, leads to the anisotropy crossover observed with increasing bilayer number. The 3-bilayer system exhibits relatively weak (002) texturing and less pronounced crystallite size anisotropy compared to higher bilayer systems. As a result, dipolar interactions and interface-induced shape anisotropy dominate, aligning the easy axis perpendicular to the column projection. With increasing bilayer number to 4 and 5, both texturing and crystallite size anisotropies become more pronounced. The competing contributions from these anisotropies nearly balance each other, resulting in the isotropic magnetic behavior observed in these intermediate systems. At 7 bilayers, the texturing and crystallite anisotropy further intensify, shifting the easy axis along the column projection. However, it is to be noted that despite the presence of strong (002) texturing and large crystallite anisotropy, especially in cobalt, a material known for its high MCA constant, the 7-bilayer system still does not exhibit a very strong uniaxial anisotropy. This indicates that the anisotropy components induced by periodic interfaces and dipolar interactions are sufficiently strong to compete with, and in some cases counteract, even significant MCA-driven uniaxial anisotropy.

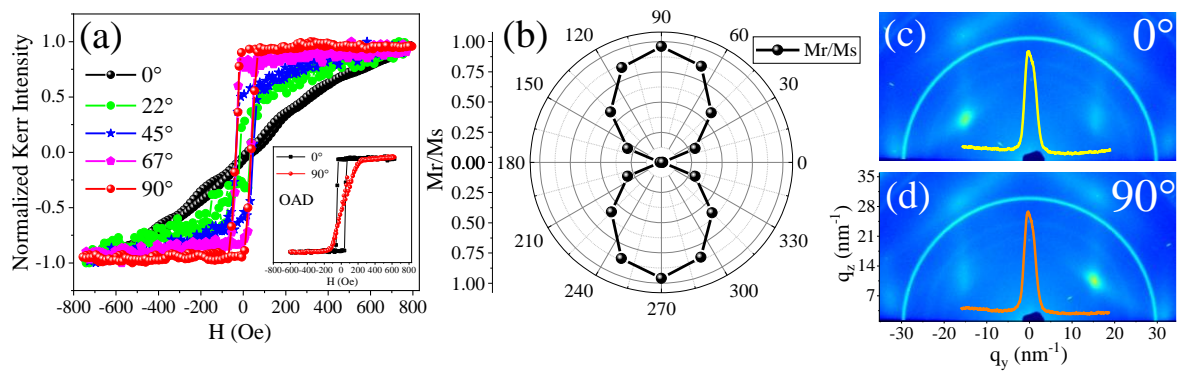

**Figure 10.** (a) shows the MOKE loops of the CFA zig-zag sample along $\phi$ = 0°, 22°, 45°, 67°, and 90°. (b) displays the polar plot of remanence magnetization variation with azimuthal angle ($\phi$ = 0°- 360°). (c, d) shows the 2DXRD images while keeping the X-ray direction along ($\phi$ = 0°) and perpendicular ($\phi$ = 90°) to the column projection.

To probe the role of MCA and further validate the emergence of the novel interface-induced anisotropy component, a 3-bilayer zigzag film of $Co_2FeAl$ (CFA) was deposited using the same structural parameters as the 3-bilayer Co film. Unlike Co, which possesses strong MCA, CFA exhibits intrinsically low MCA [21]. Interestingly, the CFA zigzag film showed a robust UMA with the easy axis oriented perpendicular to the column projection, strongly supporting the presence of interface-induced shape anisotropy in these zigzag structures. The 2DXRD patterns of the CFA zigzag film [Fig. 10 (c, d)] taken along orthogonal directions show an arc corresponding to the (220) plane with uniform intensity. Azimuthal integration (not shown) confirms no directional variation in the intensity, indicating a lack of any texturing. Moreover, due to its cubic symmetry, the crystallite size remains nearly isotropic along both directions. The fully closed hysteresis loop along the projection of columns further suggests the absence of any anisotropy component in that direction. For comparison, a conventional OAD CFA film of identical thickness was also fabricated. As expected, it exhibits UMA along the projection of columns, consistent with shape anisotropy in tilted columns. Notably, the hard-axis loop in the conventional OAD film saturates around ~200 Oe, whereas the hard-axis loop of the CFA zigzag film remains unsaturated even up to ~800 Oe, indicating a substantially stronger UMA in the latter. While both OAD and S-OAD rely on shape anisotropy to induce magnetic anisotropy, the enhancement observed in the zigzag films arises from differences in geometric aspect ratio. In conventional OAD, both the length and width of tilted columns are on the nanometer scale, whereas in S-OAD, the periodic interfaces extend laterally over microns while retaining nanometer-scale widths. This enhanced aspect ratio significantly modifies the associated demagnetization fields [5,54,55], resulting in a much stronger UMA compared to conventional OAD films.

The influence of interface-induced shape anisotropy is further evident in the magnetization behavior of the single bilayer zigzag Co film. MOKE measurements reveal a uniaxial anisotropy with the easy axis aligned along the column projection. In this case, the relatively thick individual layers (~25 nm) provide sufficient shape anisotropy to favor magnetization along the column direction. However, the presence of only a single interface result in a reduced overall anisotropy strength. Due to this interface-induced shape anisotropy component, the loop in the perpendicular direction is still wide open. On the other hand, the conventional OAD sample showed a strong UMA, with well-defined easy and hard axes along and perpendicular to the column projection, respectively. The absence of interfaces in this sample allows the shape anisotropy from the continuous columns to dominate, as expected for films deposited by OAD. These results underscore the critical role of engineered interfaces in zigzag nanostructures, offering a precise means to tune magnetic anisotropy by controlling the number of bilayers and individual layer thicknesses.

## Conclusions

This study demonstrates that S-OAD induced zigzag nanostructuring, provides an effective route to tune magnetic anisotropy in thin films. By systematically varying the number of bilayers in Co, we observe a distinct anisotropy crossover driven by competing contributions from MCA, dipolar interactions, and a novel interface-induced shape anisotropy. Synchrotron-based GISAXS and 2DXRD analyses confirm the correlation between structural evolution, such as increasing (002) texturing, anisotropic crystallite growth, and periodic interface formation, and the observed magnetic behavior. This interface-induced component, arising from periodic high-density junctions between oppositely tilted columns, proves capable of modulating or even counteracting the strong MCA of cobalt. The robust UMA in CFA zigzag films in the absence of crystallographic texture reinforces the dominance of interface effects. Compared to conventional OAD, zigzag films exhibit stronger UMA, enhanced structural stability, and prevent column merging even at higher thicknesses. This work establishes interface engineering and columnar nanostructuring within zigzag nanostructures as a versatile tool for customizing magnetic properties in functional thin films. Importantly, this study opens a plethora of new experimental avenues for probing nanoscale magnetic interactions and designing advanced thin film materials for spintronic, sensing, and data storage applications.

## Acknowledgements

The author gratefully acknowledges DESY (Hamburg, Germany), a member of the Helmholtz Association (HGF), for providing access to experimental facilities. Financial support from the India-DESY collaboration for conducting experiments at the P03 beamline of PETRA III is also sincerely acknowledged.